\documentclass[aps,prd,a4paper,amsmath,amssymb,nofootinbib,showpacs,showkeys,preprintnumbers,notitlepage,superscriptaddress,twocolumn]{revtex4-1}

\usepackage{bm}
\usepackage[dvips]{color,graphicx}

\begin{document}
\title{Dressed Polyakov loop and phase diagram of hot quark matter under magnetic field}

\author{Raoul Gatto}\email{raoul.gatto@unige.ch}
\affiliation{Departement de Physique Theorique,
 Universite de Geneve, CH-1211 Geneve 4, Switzerland}

\author{Marco Ruggieri}\email{ruggieri@yukawa.kyoto-u.ac.jp}
\affiliation{Yukawa Institute for Theoretical Physics,
 Kyoto University, Kyoto 606-8502, Japan}

\begin{abstract}
We evaluate the dressed Polyakov loop for hot quark matter in
strong magnetic field. To compute the finite temperature effective
potential, we use the Polyakov extended Nambu-Jona Lasinio model
with eight-quark interactions taken into account. The bare quark
mass is adjusted in order to reproduce the physical value of the
vacuum pion mass. Our results show that the dressed Polyakov loop
is very sensitive to the strenght of the magnetic field, and it is
capable to capture both the deconfinement crossover and the chiral
crossover. Besides, we compute self-consistently the phase diagram
of the model. We find a tiny split of the two aforementioned
crossovers as the strength of the magnetic field is increased.
Concretely, for the largest value of magnetic field investigated
here, $eB=19 m_\pi^2$, the split is of the order of $10\%$. A
qualitative comparison with other effective models and recent
Lattice results is also performed.
\end{abstract}
\pacs{12.38.Aw,12.38.Mh}\keywords{Hot quark matter, Dressed
Polyakov loop, Effective models of QCD, Deconfinement and chiral
symmetry restoration in magnetic field} \preprint{YITP-10-55}
\maketitle

\section{Introduction}
The nature of the Quantum Chromodynamics (QCD) vacuum is one of
the most intriguing aspects of modern physics. Besides, it is very
hard to get a full understanding of its properties, because its
most important characteristics, namely chiral symmetry breaking
and color confinement, have a non-perturbative origin, and the use
of perturbative methods is useless. One of the best strategies to
overcome this problem is offered by Lattice QCD simulations at
zero chemical potential
(see~\cite{deForcrand:2006pv,Aoki:2006br,Bazavov:2009zn,Cheng:2009be}
for several examples and see also references therein).  At
vanishing quark chemical potential, it is almost established that
two crossovers take place at nearly the same temperature; one for
quark deconfinement, and another one for the (approximate)
restoration of chiral symmetry.  It is still under debate whether
two crossovers should occur at exactly the same temperature, see
for example the report in Ref.~\cite{Aoki:2006br}.

An alternative approach to the physics of strong interactions,
which is capable to capture some of the non-perturbative
properties of the QCD vacuum, at the same time being easy to
manage mathematically, is the Nambu-Jona Lasinio (NJL)
model~\cite{Nambu:1961tp}, see also Refs.~\cite{revNJL} for
reviews. In this model, the QCD gluon-mediated interactions are
replaced by effective interactions among quarks, which are built
in order to respect the global symmetries of QCD. Since dynamical
gluons are absent in this model, it is not a gauge theory.
However, it shares the global symmetries of the QCD action;
moreover, the parameters of the NJL model are fixed to reproduce
some phenomenological quantity of the QCD vacuum: in its simplest
version, the pion decay constant, the vacuum pion mass and the
vacuum chiral condensate are reproduced. Therefore, it is
reasonable that the main characteristics of its phase diagram
represent, at least qualitatively, those of QCD.

Critically speaking, the worst aspect of the NJL model is that it
lacks confinement: massive quark poles of the quark propagator are
present at any temperature and/or chemical potential. It is well
known that color confinement can be described in terms of the
center symmetry of the color gauge group and of the Polyakov
loop~\cite{Polyakovetal}, which is an order parameter for the
center symmetry. Motivated by this property, the Polyakov extended
Nambu-Jona Lasinio model (P-NJL model) has been
introduced~\cite{Meisinger:1995ih,Fukushima:2003fw}, in which the
concept of statistical confinement replaces that of the true
confinement of QCD, and an effective interaction among the chiral
condensate and the Polyakov loop is achieved by a covariant
coupling of quarks with a background temporal gluon field. In the
literature, there are several studies about various aspects of the
P-NJL model. Its phase structure with two flavors and symmetric
quark matter has been investigated in
Refs.~\cite{Ratti:2005jh,Roessner:2006xn,Megias:2006bn,Sasaki:2006ww};
PNJL model with a Van der Monde term has been considered
in~\cite{Ghosh:2007wy}; phase structure with 2+1 flavors has been
studied in Refs.~\cite{Fukushima:2008wg}; possible realization of
the quarkyonic phase~\cite{McLerran:2007qj} has been discussed
in~\cite{Fukushima:2008wg,Abuki:2008nm}; mass dependence of the
phase diagram, and a possible emergence of the quarkyonic phase,
is investigated in~\cite{Kahara:2010wh}; phase diagram with
imaginary chemical potential has been studied
in~\cite{Sakai:2008py,Sakai:2009dv}; dual quark condensate has
been computed in~\cite{Kashiwa:2009ki}; neutral phases have been
investigated in~\cite{Abuki:2008tx}; phase diagram with asymmetric
quark matter have been studied in~\cite{Sasaki:2010jz}; non-local
extension has been introduced in~\cite{Hell:2008cc}; role of
eight-quark interactions in the PNJL context has been elucidated
in~\cite{Kashiwa:2007hw}.

The modification of the QCD vacuum, and of its thermal excitations
as well, under the influence of external fields, is an attractive
topic. Firstly, it is extremely interesting to understand how an
external field can modify the main characteristics of confinement
and spontaneous chiral symmetry breaking. Lattice studies on the
response to external magnetic fields can be found
in~\cite{D'Elia:2010nq,Buividovich:2009my,Buividovich:2008wf}. QCD
in chromo-magnetic fields has been investigated on the Lattice
in~\cite{Cea:2002wx,Cea:2007yv}. Previous studies of QCD in
magnetic fields, and of QCD-like theories as well, can be found in
Refs.~\cite{Klevansky:1989vi,Gusynin:1995nb,Klimenko:1990rh,Agasian:2008tb}.
A self-consistent model calculations of magnetic catalysis and of
deconfinement pseudo-critical temperature in magnetic field, has
been performed firstly in~\cite{Fukushima:2010fe} within the PNJL
model, and then in~\cite{Mizher:2010zb} using the Polyakov
extended quark-meson model. Effective models in chromo-magnetic
fields have been considered in~\cite{Campanelli:2009sc}. Besides,
strong magnetic fields might be produced in non-central heavy ion
collisions~\cite{Kharzeev:2007jp,Skokov:2009qp}. In this case, it
has been argued that the non-trivial topological structure of
thermal QCD gives rise to Chiral Magnetic Effect (CME)
\cite{Kharzeev:2007jp,Buividovich:2009wi,Fukushima:2008xe}.

Beside the Polyakov loop, it has been
suggested~\cite{Bilgici:2008qy} that another observable which is
an order parameter for the center symmetry, hence for confinement,
is the dressed Polyakov loop. From the mathematical point of view,
the dressed Polyakov loop is built from the canonical (called {\em
thin}) Polyakov loop, by dressing it with higher order loops,
which wind once around the compact temporal direction. In this
context, the order of a loop is given by its length; the thin
Polyakov loop corresponds to the shortest one. The dressing
becomes more important when quark masses are finite (Polyakov loop
is an exact order parameter for confinement-deconfinement only in
the ideal case of static quarks with infinite masses). In
Refs.~\cite{Fischer:2009wc} the dressed Polyakov loop has been
computed within the scheme of truncated Schwinger-Dyson equations,
with a model for resummed quark-gluon vertex and in-medium gluon
propagator computed on the Lattice. Within the Nambu-Jona Lasinio
model, in which the QCD interaction among quarks is replaced by a
contact four-fermion interaction, $\Sigma_1$ has been computed at
finite temperature and chemical potential
in~\cite{Mukherjee:2010cp}. Finally, the dressed Polyakov loop has
been computed within the PNJL model in~\cite{Kashiwa:2009ki} at
finite temperature.

In this article, we compute the phase structure and the dressed
Polyakov loop of hot quark matter at zero chemical potential, in
an external magnetic field. To compute the effective potential, we
rely on the PNJL model of strongly interacting quarks. We will
limit ourselves to the one-loop approximation (saddle point),
which is enough to draw a phase structure. The novelty of the
present article is manyfold. Firstly, we introduce the eight-quark
interaction~\cite{Osipov:2006ev,Kashiwa:2006rc,Osipov:2005tq,Osipov:2007je}
in the PNJL model in an external magnetic field (previous studies
of the PNJL model in magnetic and chromomagnetic fields neglected
this kind of interaction). Within the Nambu-Jona Lasinio model, it
has been shown that the eight-quark interactions naturally lowers
the pseudo-critical temperature for (approximate) chiral symmetry
restoration. Magnetic catalysis in the NJL model with multi-quark
interactions has been investigated in~\cite{Osipov:2007je}.
However, in those studies, the computation of quantities relevant
for deconfinement crossover is lacking. On the other hand,
in~\cite{Kashiwa:2007hw} the Polyakov extended NJL model with
multi-quark interaction has been investigated, but without
magnetic field. It is of interest, then, to study the response of
quark matter to magnetic fields in the framework of the PNJL model
with eight-quark interaction. In doing this, we will consider
quarks with finite values of bare mass, fixed to reproduce the
vacuum pion mass, while in a previous
study~\cite{Fukushima:2010fe} this problem was studied only in the
chiral limit.

Moreover, we compute the dressed Polyakov loop, $\Sigma_1$, in a
magnetic field. Along this line, we anticipate one of our results,
namely that the dressed Polyakov loop, $\Sigma_1$, is capable to
feel both the Polyakov loop and the chiral condensate crossovers,
whatever the strength of the magnetic field is. This occurs
despite the tiny split of the two crossovers, which we observe at
sufficiently strong magnetic field strength. Therefore, in view of
an effective theory for finite temperature QCD in terms of just
one order parameter, our results are encouraging.

The plan of the paper is as follows. In Section II, we present the
model we use. In Section III, we show and discuss our numerical
results. Finally, in Section IV, we draw our conclusions.

\section{Dressed Polyakov loop in the effective model}
In this article, we model quark matter by the following Lagrangian
density
\begin{eqnarray}
{\cal L} &=& \bar q\left(i\gamma^\mu D_\mu - m_0\right)q
        + g_\sigma\left[(\bar q q)^2 + (\bar q i \gamma_5 \bm\tau q)^2\right] \nonumber \\
        &&+g_8\left[(\bar q q)^2 + (\bar q i \gamma_5 \bm\tau
        q)^2\right]^2~,
\label{eq:lagr}
\end{eqnarray}
which corresponds to the NJL lagrangian with multi-quark
interactions~\cite{Osipov:2006ev}. The covariant derivative embeds
the quark coupling to the external magnetic field and to the
background gluon field as well, as we will see explicitly below.
In Eq.~\eqref{eq:lagr}, $q$ represents a quark field in the
fundamental representation of color and flavor (indices are
suppressed for notational simplicity); $m_0$ is the bare quark
mass, which is fixed to reproduce the pion mass in the vacuum,
$m_\pi = 139$ MeV. Our interaction in Eq.~\eqref{eq:lagr} consists
of a four-quark term, whose coupling $g_\sigma$ has inverse mass
dimension two, and an eight-quark term, whose coupling constant
$g_8$ has inverse mass dimension eight.

\begin{widetext}
The evaluation of the bulk thermodynamic quantities requires we
compute the quantum effective action of the model. This cannot be
done exactly. Hence, we rely ourselves to the one-loop
approximation for the partition function, which amounts to take
the classical contribution plus the fermion determinant. The
one-loop thermodynamic potential of quark matter in external
fields has been discussed
in~\cite{Fukushima:2010fe,Campanelli:2009sc}, in the case of
canonical antiperiodic boundary conditions;
following~\cite{Kashiwa:2009ki}, it is easy to generalize it to
the more general case of twisted boundary conditions:
\begin{eqnarray}
\Omega &=& {\cal U}(P,\bar P, T) + \frac{\sigma^2}{g_\sigma} +
\frac{3\sigma^4 g_8}{g_\sigma^4}
        - \sum_{f=u,d}\frac{|q_f e B|}{2\pi}
           \sum_{k}\alpha_{k}\int_{-\infty}^{+\infty}\frac{dp_z}{2\pi}g_\Lambda(p_z,k)\omega_{k}(p_z)
  \nonumber \\
      &-&T\sum_{f=u,d}\frac{|q_f e B|}{2\pi}\sum_{k}\alpha_{k}
        \int_{-\infty}^{+\infty}\frac{dp_z}{2\pi}
        \log\left(1+3P e^{-\beta\cal{E}_-} + 3\bar{P}e^{-2\beta\cal{E}_-}+e^{-3\beta\cal{E}_-} \right)
      \nonumber \\
     &-&T\sum_{f=u,d}\frac{|q_f e B|}{2\pi}\sum_{k}\alpha_{k}
        \int_{-\infty}^{+\infty}\frac{dp_z}{2\pi}
        \log\left(1+3\bar P e^{-\beta\cal{E}_+} + 3Pe^{-2\beta\cal{E}_+}+e^{-3\beta\cal{E}_+} \right)~.
\label{eq:Om1}
\end{eqnarray}
\end{widetext}
In the previous equation, $\sigma=g_\sigma\langle\bar q q\rangle = 2g_\sigma\langle\bar u u\rangle$;
$k$ is a non-negative integer which labels the Landau level; $\alpha_k = \delta_{k0} + 2(1-\delta_{k0})$ counts the degeneracy of the $k-$th Landau level.
We have put
\begin{equation}
\omega_k(p_z)^2=p_z^2 + 2|q_f eB| k + M^2~,
\end{equation}
with $M = m_0 - 2\sigma -4\sigma^3 g_8/g_\sigma^3$~. The arguments of the thermal exponentials are
defined as
\begin{equation}
{\cal E}_\pm = \omega_k(p_z) \pm \frac{i(\varphi-\pi)}{\beta}~,
\end{equation}
with $\varphi$ defined in Eq.~\eqref{eq:phi}.

The vacuum part of the thermodynamic potential, $\Omega(T=0)$, is
ultraviolet divergent.  This divergence is transmitted to the
self-consistent equations which determine the chiral condensate
and the expectation value of the Polyakov loop. In this article,
we use a smooth regularization procedure by introducing a form
factor $g_\Lambda(p)$ in the diverging zero-point energy. Our
choice of $g_\Lambda(p)$ is
\begin{equation}
 g_\Lambda(p) = \frac{\Lambda^{2N}}
  {\Lambda^{2N} + (p_z^2 + 2|q_f eB| k)^{N}}~;
  \label{eq:UVr}
\end{equation}
we choose two values of $N$, namely $N=5$ and $N=7$.

The potential term $\mathcal{U}[P,\bar P,T]$ in Eq.~\eqref{eq:Om1}
is built by hand in order to reproduce the pure gluonic lattice
data~\cite{Ratti:2005jh}.  Among several different potential
choices~\cite{Schaefer:2009ui} we adopt the following logarithmic
form~\cite{Fukushima:2003fw,Ratti:2005jh},
\begin{equation}
 \begin{split}
 & \mathcal{U}[P,\bar P,T] = T^4\biggl\{-\frac{a(T)}{2}
  \bar\Phi \Phi \\
 &\qquad + b(T)\ln\bigl[ 1-6\bar PP + 4(\bar P^3 + P^3)
  -3(\bar PP)^2 \bigr] \biggr\} \;,
 \end{split}
\label{eq:Poly}
\end{equation}
with three model parameters (one of four is constrained by the
Stefan-Boltzmann limit),
\begin{equation}
 \begin{split}
 a(T) &= a_0 + a_1 \left(\frac{T_0}{T}\right)
 + a_2 \left(\frac{T_0}{T}\right)^2 , \\
 b(T) &= b_3\left(\frac{T_0}{T}\right)^3 \;.
 \end{split}
\label{eq:lp}
\end{equation}
The standard choice of the parameters reads~\cite{Ratti:2005jh};
\begin{equation}
 a_0 = 3.51\,, \quad a_1 = -2.47\,, \quad
 a_2 = 15.2\,, \quad b_3 = -1.75\,.
\end{equation}
The parameter $T_0$ in Eq.~\eqref{eq:Poly} sets the deconfinement
scale in the pure gauge theory, i.e.\ $T_c = 270$ \text{MeV}.

Following~\cite{Bilgici:2008qy}, in order to define the dressed
Polyakov loop, we work in a finite Euclidean volume with
temperature extension $\beta=1/T$. We take twisted fermion
boundary conditions along the compact temporal direction,
\begin{equation}
q(\bm x,\beta) = e^{-i\varphi}q(\bm x,0)~,~~~\varphi\in[0,2\pi]~,
\label{eq:phi}
\end{equation}
while for spatial directions the usual periodic boundary condition
is taken. The canonical antiperiodic boundary condition for the
quantization of fermions at finite temperature, is obtained by
taking $\varphi = \pi$ in the previous equation. The dual quark
condensate, $\tilde\Sigma_n$, is defined as
\begin{equation}
\tilde\Sigma_n(m,V) =
\int_0^{2\pi}\frac{d\varphi}{2\pi}\frac{e^{-i\varphi n}}{V}
\langle\bar q q\rangle_G~, \label{eq:Sn0}
\end{equation}
where $n$ is an integer. The expectation value
$\langle\bm\cdot\rangle_G$ denotes the path integral over gauge
field configurations. An important point is that in the
computation of the expectation value, the twisted boundary
conditions acts only on the fermion determinant; the gauge fields
are taken to be quantized with the canonical periodic boundary
condition.

Using a lattice regularization, it has been shown
in~\cite{Bilgici:2008qy} that Eq.~\eqref{eq:Sn0} can be expanded
in terms of loops which wind $n$ times along the compact time
direction. In particular, the case $n=1$ is called the dressed
Polyakov loop; it corresponds to a sum of loops winding just once
along the time direction. These correspond to the thin Polyakov
loop (the loop with shortest length) plus higher order loops, the
order being proportional to the length of the loop. Each higher
order loop is weighed by an inverse power of the quark mass.
Because of the weight, in the infinite quark mass limit only the
thin Polyakov loop survives; for this reason, the dressed Polyakov
loop can be viewed as a mathematical dressing of the thin loop, by
virtue of longer loops, the latter being more and more important
as the quark mass tends to smaller values.

If we denote by $z$ an element of the center of the color gauge
group, then it is easy to show that $\tilde\Sigma_n \rightarrow
z^n\tilde\Sigma_n$. It then follows that, under the center of the
symmetry group $Z_3$, the dressed Polyakov loop is an order
parameter for the center symmetry, with the same transformation
rule of the thin Polyakov loop. Since the center symmetry is
spontaneously broken in the deconfinement phase and restored in
the confinement phase~\cite{Polyakovetal} (in presence of
dynamical quarks, it is only approximately restored), the dressed
Polyakov loop can be regarded as an order parameter for the
confinement-deconfinement transition as well.

For later convenience, we scale the definition of the dressed
Polyakov loop in Eq.~\eqref{eq:Sn0}, and introduce
\begin{eqnarray}
\Sigma_1 &=& -2\pi
g_\sigma\int_0^{2\pi}\frac{d\varphi}{2\pi}e^{-i\varphi}
\langle\bar q q\rangle_G~, \nonumber \\
 &=& -
\int_0^{2\pi}\!d\varphi~e^{-i\varphi}
\sigma(\varphi)~,\label{eq:SnPNJL}
\end{eqnarray}
where $\sigma(\varphi)$ corresponds to the expectation value of
the $\sigma$ field computed keeping twisted boundary conditions
for fermions.

\section{Numerical results}
In this Section, we show our results. The main goal to achieve
numerically is the solution of the gap equations,
\begin{equation}
\frac{\partial\Omega}{\partial\sigma}=0~,~~~
\frac{\partial\Omega}{\partial P}=0~. \label{eq:GE}
\end{equation}
This is done by using a globally convergent algorithm with
backtrack~\cite{NumericalRecipes}. From the very definition of the
dressed Polyakov loop, Eq.~\eqref{eq:Sn0}, the twisted boundary
condition, Eq.~\eqref{eq:phi}, must be imposed only in
$D_\varphi$. Therefore, we firstly compute the expectation value
of the Polyakov loop and to the chiral condensate, taking
$\varphi=\pi$. Then, in order to compute the dressed Polyakov loop
using Eq.~\eqref{eq:SnPNJL}, we compute the $\varphi$-dependent
chiral condensate using the first of Eq.~\eqref{eq:GE}, keeping
the expectation value of the Polyakov loop fixed at its value at
$\varphi=\pi$~\cite{Kashiwa:2009ki}.

\begin{table}[t!]
\caption{\label{Tab:para}Parameters of the model for the two
choices of the UV-regulator.}
\begin{ruledtabular}
\begin{tabular}{ccccc}
 &$\Lambda$ (MeV)& $m_0$ (MeV) & $g_\sigma$ (MeV)$^{-2}$ & $g_8$ (MeV)$^{-8}$\\
\hline

$N=5$ &$588.657$ & 5.61& $5\times 10^{-6}$ & $6\times 10^{-22}$\\
\hline

$N=7$ & $603.475$ &5.61 & $4.92\times 10^{-6}$ & $6.8\times
10^{-22}$
\end{tabular}
\end{ruledtabular}
\end{table}

In this study, we report results obtained using the UV-regulator
specified in Eq.~\eqref{eq:UVr} with $N=5$ and $N=7$. As expected,
there is no qualitative difference among the pictures that the two
regularization schemes lead to. As a consequence, concrete results
are shown only for the case $N=5$; for what concerns the case
$N=7$, we collect the pseudo-critical temperatures in
Table~\ref{Tab:fit}. We have also checked that the results are
qualitatively unchanged if we use a hard cutoff scheme instead of
the smooth UV-regulator. The parameter set for both cases is
specified in Table~\ref{Tab:para}. In the case $N=5$, they are
obtained by the requirements that the vacuum pion mass is $m_\pi =
139$ MeV, the pion decay constant $f_\pi = 92.4$ MeV and the
vacuum chiral condensate $\langle\bar u u\rangle
\approx(-241~\text{MeV})^3$. In this case, the chiral and
deconfinement pseudo-critical temperatures at zero magnetic field
are $T_0^\chi = T_0^P = 175$~\text{MeV}. Similarly, for the case
$N=7$, the chiral and deconfinement pseudo-critical temperatures
at zero magnetic field are $T_0^\chi = 176$~\text{MeV} and $T_0^P
= 175$~\text{MeV}, respectively; the zero temperature chiral
condensate at zero magnetic field strength is fixed to
$\langle\bar u u\rangle \approx (-246~\text{MeV})^3$.

We remark that the main effect of the eight-quark interaction in
Eq.~\eqref{eq:lagr} is to lower the pseudo-critical temperature of
the crossovers. This has been already discussed several times in
the literature~\cite{Kashiwa:2006rc,Osipov:2006ev}, in the context
of both the NJL and the PNJL models. Therefore, it is not
necessary to discuss it further here, while at the same time we
prefer to stress the results that have not been discussed yet.

In order to identify the pseudo-critical temperatures, we have
define the {\em effective susceptibilities} as
\begin{equation}
\chi_A = (m_\pi)^g\left|\frac{d A}{d
T}\right|~,~~~A=\sigma,P,\Sigma_1~. \label{eq:efs}
\end{equation}
Strictly speaking, the quantities defined in the previous equation
are not true susceptibilities. Nevertheless, they allow to
represent faithfully the pseudo-critical region, that is, the
range in temperature in which the various crossovers take place.
Therefore, for our purposes it is enough to compute these
quantities. In Equation~\eqref{eq:efs}, the appropriate power of
$m_\pi$ is introduced just for a matter of convenience, in order
to have a dimensionless quantity; therefore, $g=0$ if
$A=\sigma,\Sigma_1$, and $g=1$ if $A=P$.

\subsection{Condensates and dressed Polyakov loop}

\begin{figure*}
\begin{center}
\includegraphics[width=9cm]{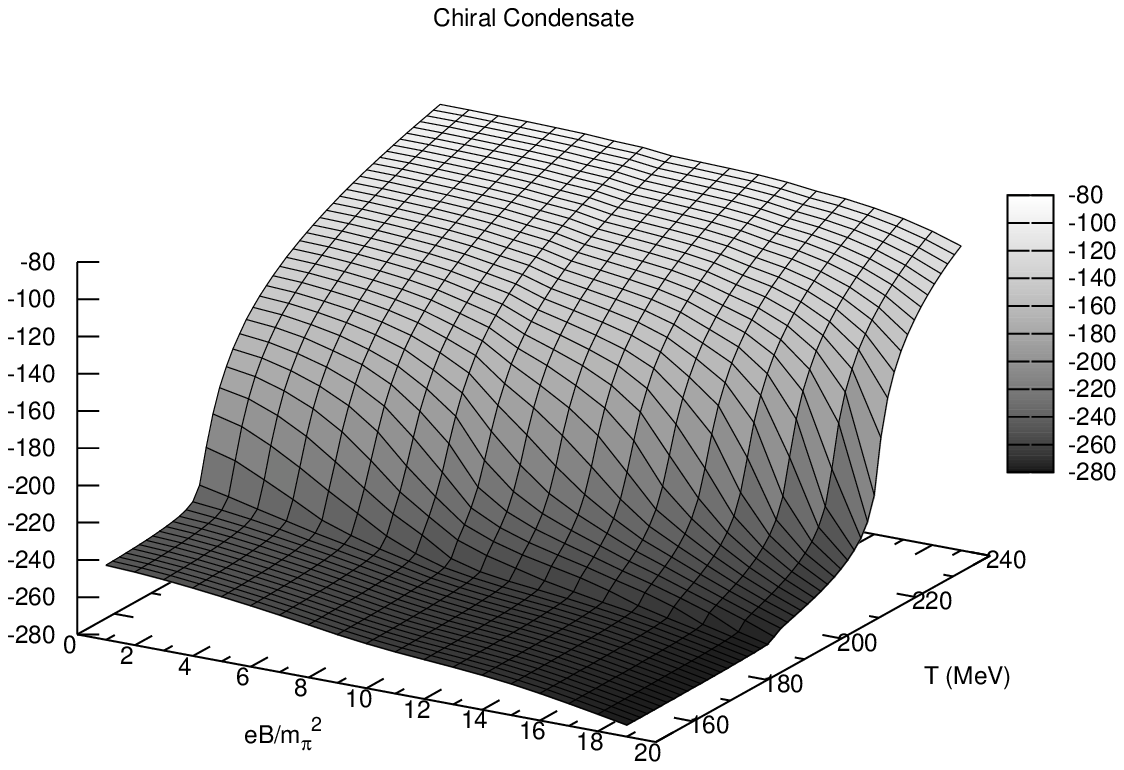}~~~\includegraphics[width=9cm]{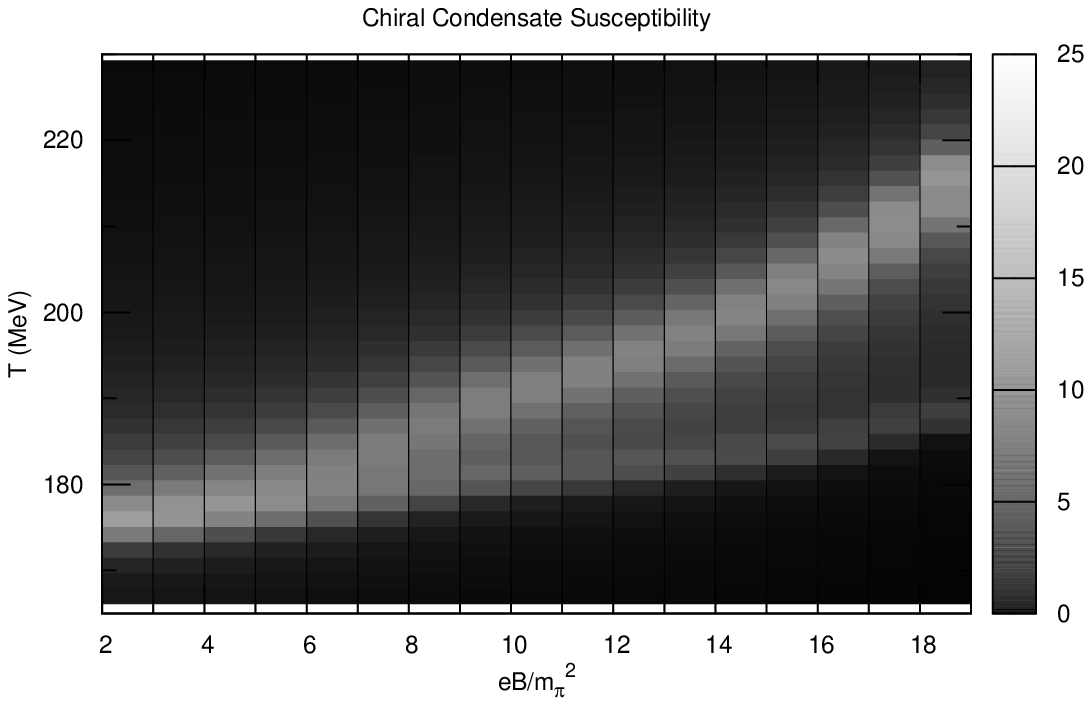}\\
\includegraphics[width=9cm]{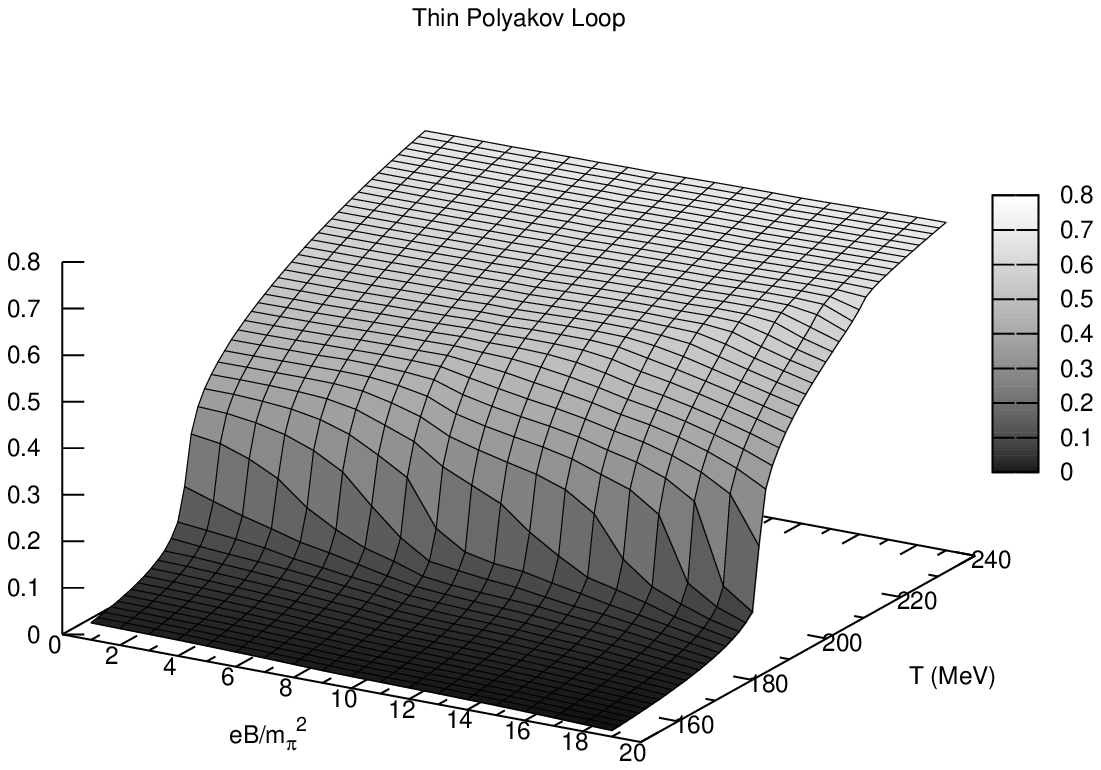}~~~\includegraphics[width=9cm]{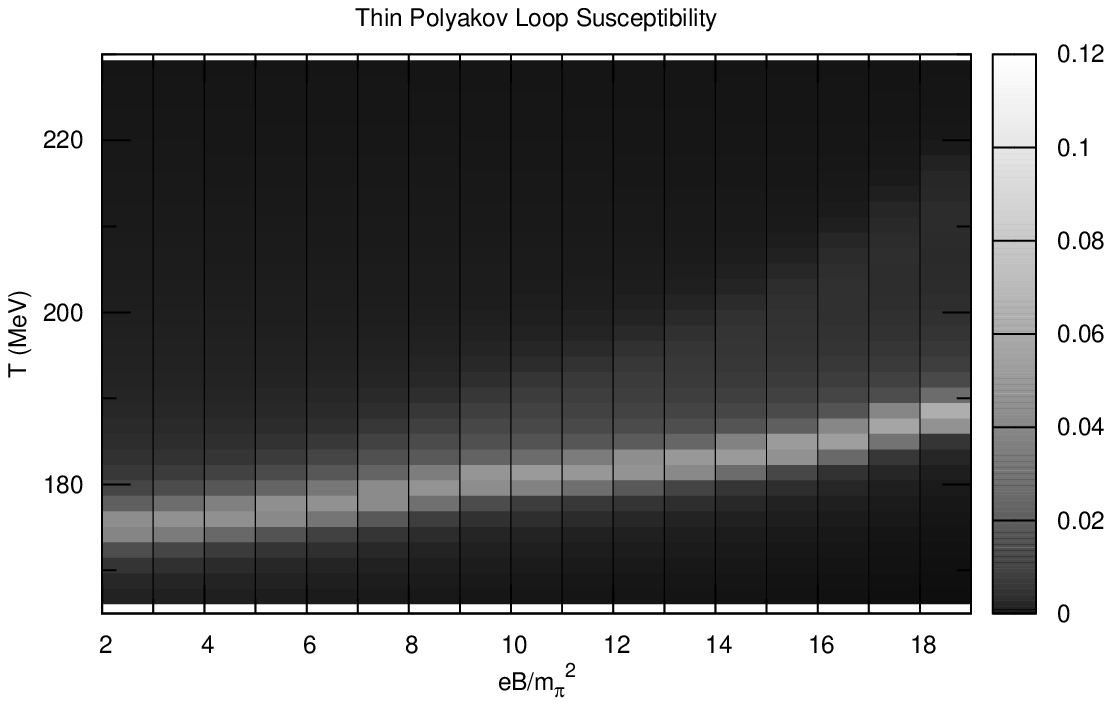}\\
\includegraphics[width=9cm]{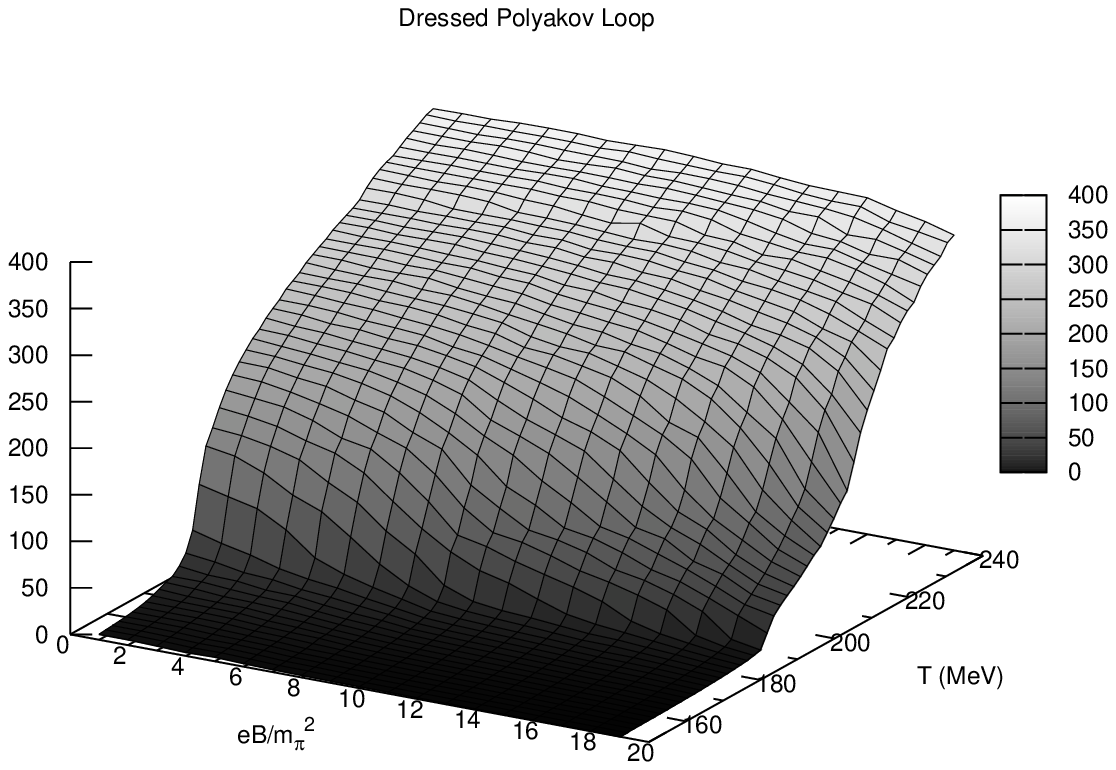}~~~\includegraphics[width=9cm]{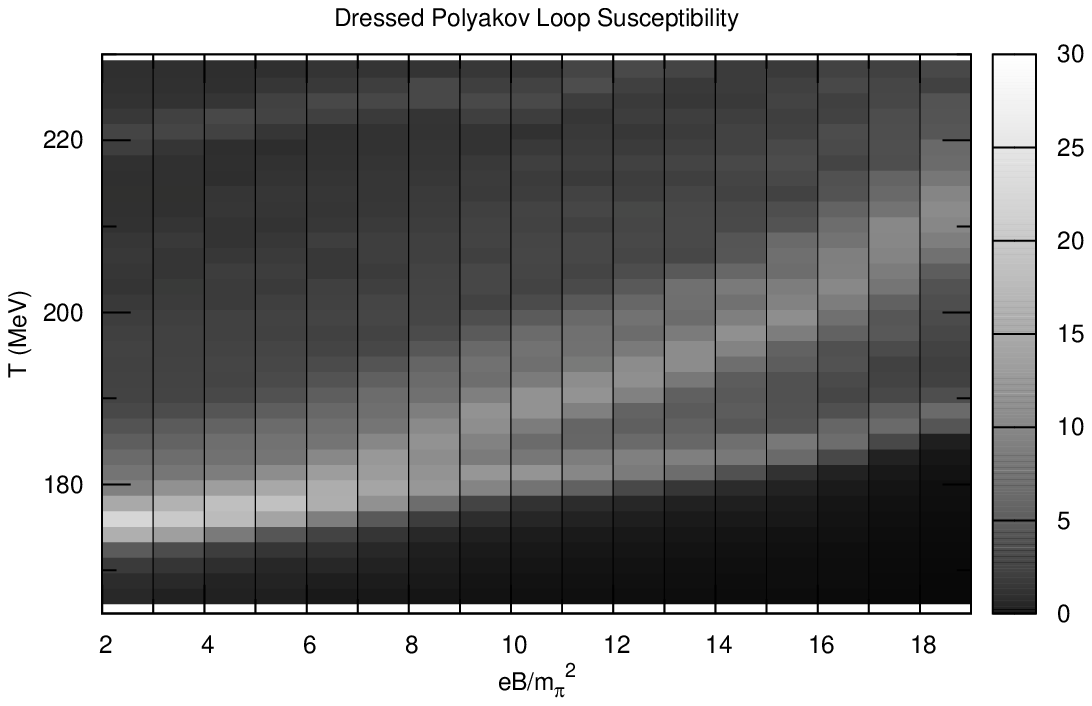}
\end{center}
\caption{\label{Fig:S3D} {\em Left panel}. Chiral condensate,
Polyakov loop and dressed Polyakov loop as a function of
temperature and magnetic field, for the case $N=5$. {\em Right
panel}. Contour plots of the raw data of the effective
susceptibilities. The lighter the color, the higher the
susceptibility. Vertical axes correspond to temperature (in MeV);
horizontal axes represent magnetic field $eB/m_\pi^2$. For the
dressed Polyakov loop susceptibility, the bifurcation of the
critical region is evident.}
\end{figure*}

From now on, we fix $N=5$ unless specified. The results for this
case are collected in the form of three-dimensional plots in
Fig.~\ref{Fig:S3D} (for the case $N=7$ the plots do not differ
qualitatively). In the left panel, we plot the chiral condensate
$\langle\bar u u\rangle^{1/3}$, the expectation value of the
Polyakov loop, and the dressed Polyakov loop $\Sigma_1$, as a
function of temperature and magnetic field. In the right panel, we
show the contour plots of the raw data of the effective
susceptibilities. The lighter the color, the higher the
susceptibility. In the contour plots, the vertical axes correspond
to temperature (measured in MeV); the horizontal axes represent
the magnetic field $eB/m_\pi^2$.

\begin{figure*}
\begin{center}
\includegraphics[width=8cm]{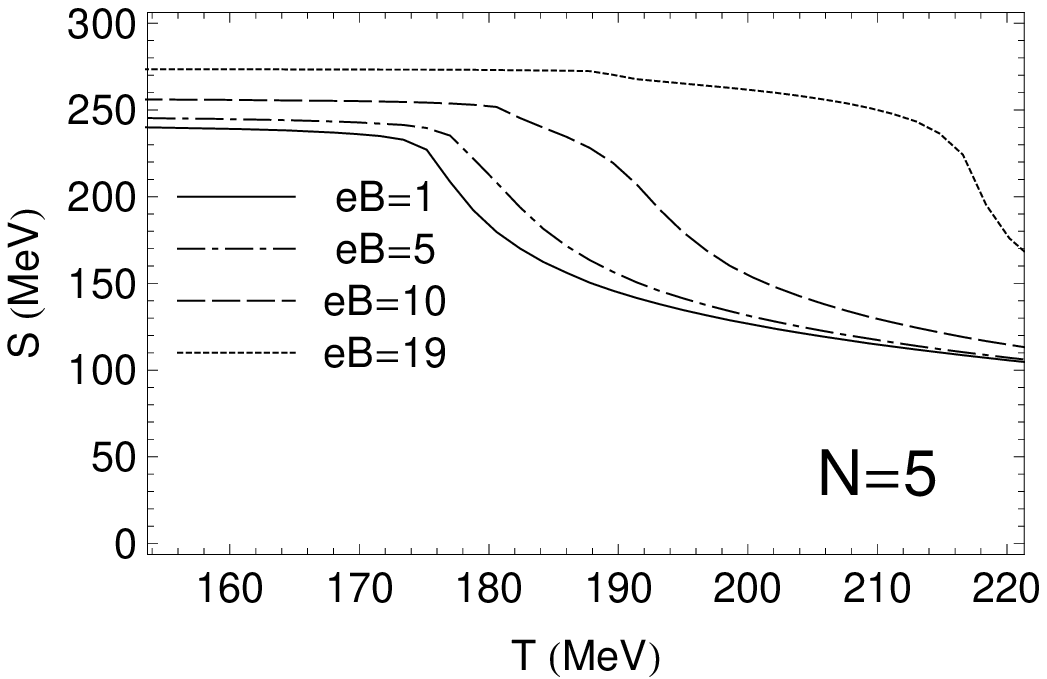}~~~\includegraphics[width=8cm]{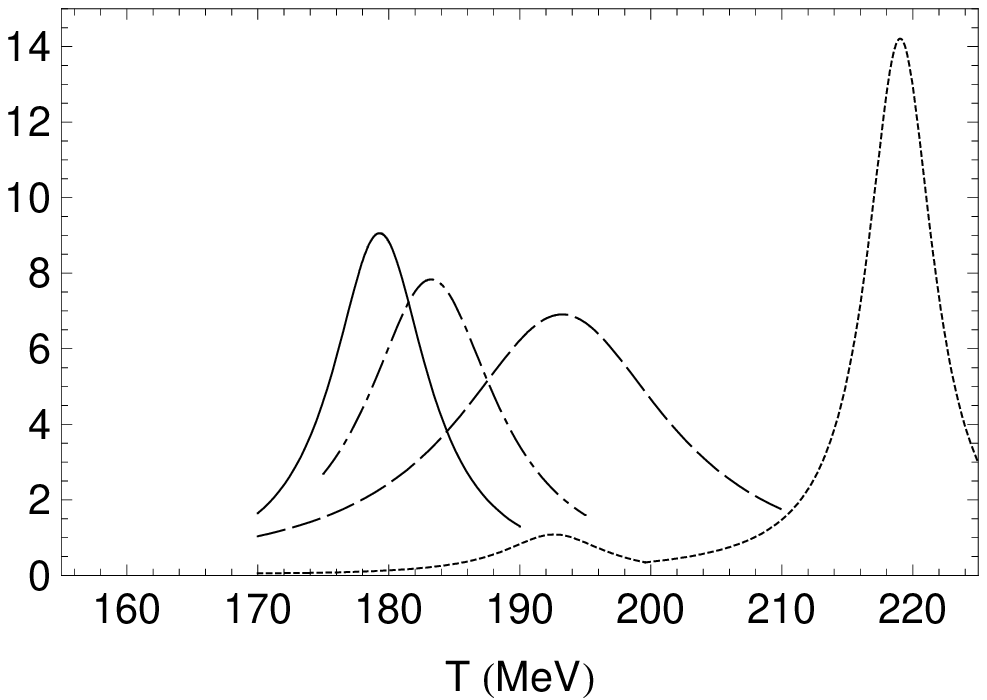}\\
\includegraphics[width=8cm]{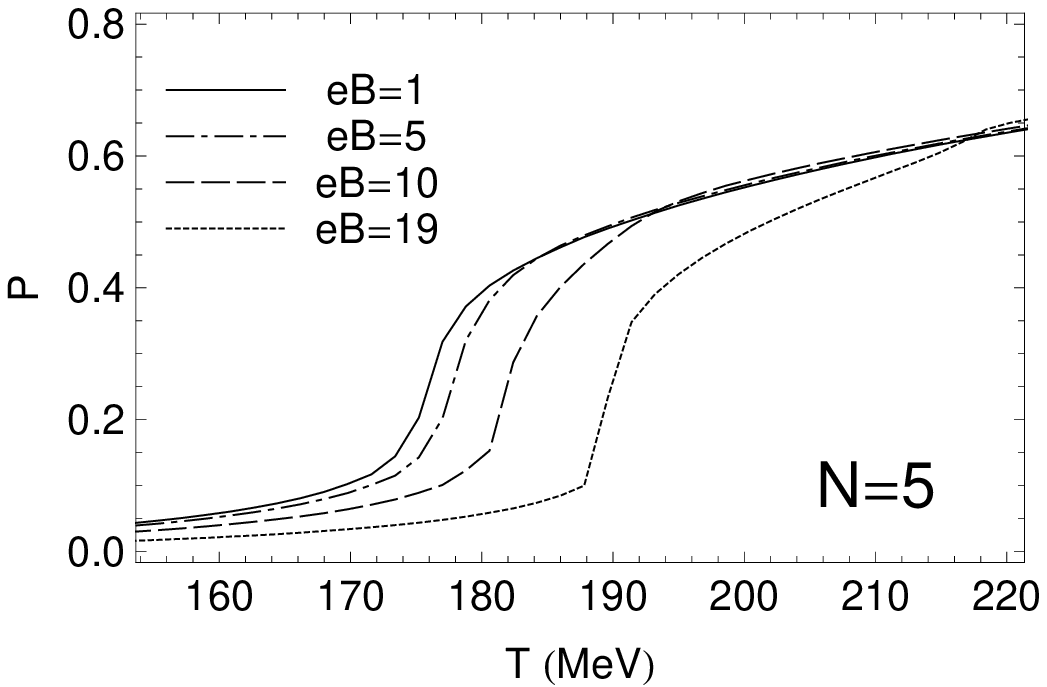}~~~\includegraphics[width=8cm]{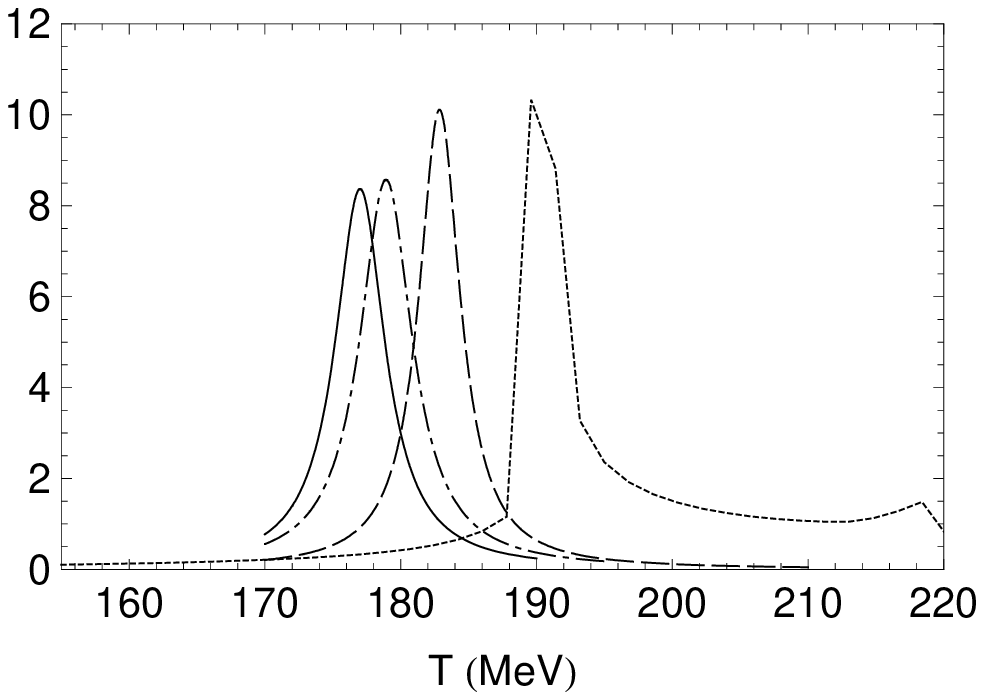}\\
\includegraphics[width=8cm]{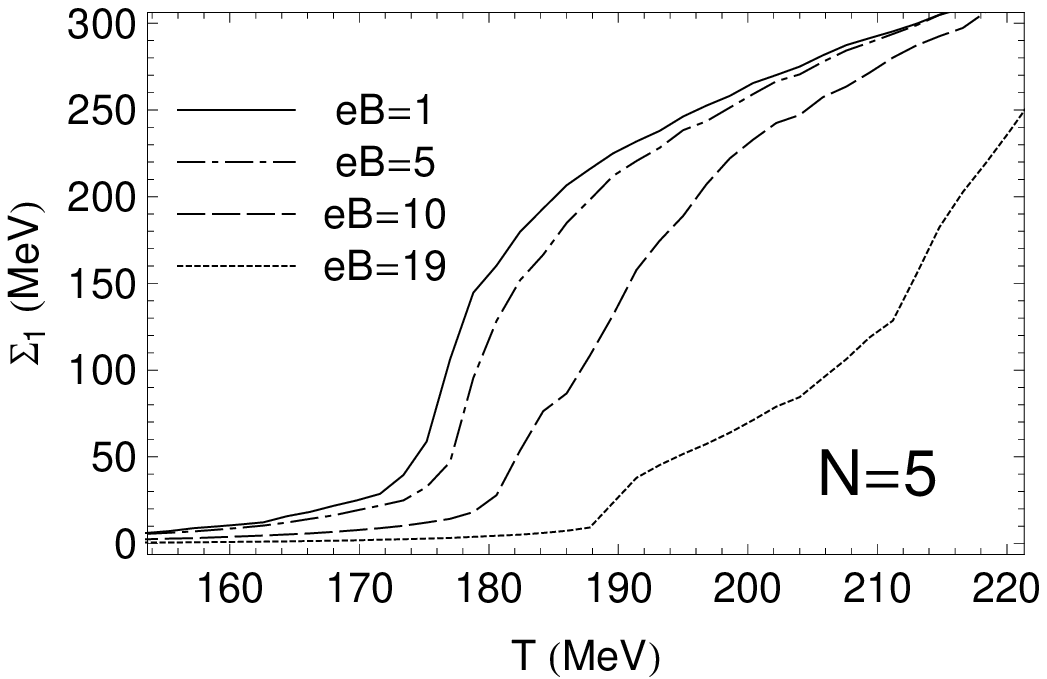}~~~\includegraphics[width=8cm]{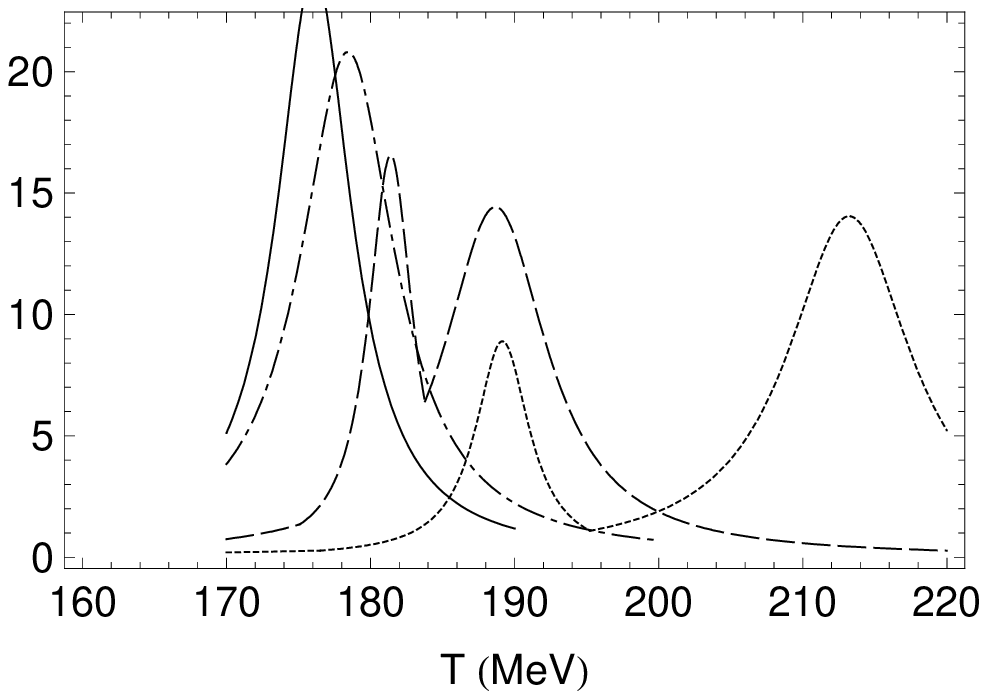}
\end{center}
\caption{\label{Fig:cN5}{\em Left panel}. Chiral condensate
$S=|\langle\bar u u\rangle|^{1/3}$ (upper panel), Polyakov loop
(middle panel) and $\Sigma_1$ (lower panel) as a function of
temperature, for several values of the applied magnetic field
strength, measured in units of $m_\pi^2$. In the figures, $N=5$
corresponds to the order of the UV-regulator in
Eq.~\eqref{eq:UVr}.  {\em Right panel}. Effective
susceptibilities, defined in Eq.~\eqref{eq:efs}, as a function of
temperature, for several values of $eB$. Conventions for lines are
the same as in the left panel.}
\end{figure*}

We slice the three dimensional plots in Fig.~\ref{Fig:S3D} at
fixed value of the magnetic field strength, and show the results
in Fig.~\ref{Fig:cN5}, where we plot the chiral condensate
$S=|\langle\bar u u\rangle|^{1/3}$ (upper panel), the Polyakov
loop (middle panel) and $\Sigma_1$ (lower panel) as a function of
temperature, for several values of the applied magnetic field
strength, measured in units of $m_\pi^2$. In the right panel, we
plot fits of the effective susceptibilities in the critical
regions, as a function of temperature. The fits are obtained from
the raw data, using Breit-Wigner-like fitting functions. The
details of the fitting procedure are not relevant for the present
discussion. For graphical reasons, in Fig.~\ref{Fig:S3D} we plot
the chiral condensate with its sign; on the other hand, in
Fig.~\ref{Fig:cN5} we take the absolute value of this quantity.

The qualitative behavior of the chiral condensate, and of the
Polyakov loop as well, is similar to that found in a previous
study within the PNJL model in the chiral
limit~\cite{Fukushima:2010fe}. Quantitatively, the main difference
with the case of the chiral limit, is that in the latter the
chiral restoration at large temperature is a true second order
phase transition (in other model calculations it has been reported
that the phase transition might become of the first order at very
large magnetic field strengths~\cite{Agasian:2008tb}). On the
other hand, in the case under investigation, chiral symmetry is
always broken explicitly because of the bare quark masses; as a
consequence, the second order phase transition is replaced by a
smooth crossover.

Another interesting aspect, observed also in the chiral
limit~\cite{Fukushima:2010fe}, is that the Polyakov loop crossover
temperature, is less sensitive to the strength of the magnetic
field than the same quantity computed for the chiral condensate.
It is useful, for illustration purpose, to quantify the net shift
of the pseudo-critical temperatures, for the largest value of
magnetic field we have studied, $eB = 19 m_\pi^2$. In this case,
if we take $N=5$ (for $N=7$ the results are similar), then the two
crossover occur simultaneously at $eB=0$, at the temperature
$T_0^\chi = T_0^P = 175$~\text{MeV}; for $eB = 19 m_\pi^2$, we
find $T_\chi = 219$~\text{MeV} and $T_P = 190$~\text{MeV}.
Therefore, the chiral crossover is shifted approximately by
$25.1\%$, to be compared with the more modest shift of the
Polyakov loop crossover, which is $\approx 8.6\%$.

The split of the two critical temperatures at a so large value of
the magnetic field strength is only of $15\%$; on the Lattice, no
split is observed~\cite{D'Elia:2010nq}, and a modest increase of
the critical temperature is measured. Therefore, we are in partial
agreement with the Lattice results, in the sense that the raising
of the critical lines is observed also in our model calculation;
for what concerns the split of the two crossovers, we can take our
${\cal O}(10\%)$ split as a consequence of the crudeness of the
model at hand. On the Lattice, the smaller pion mass used is of
the order of $200$ MeV~\cite{D'Elia:2010nq}. We have verified that
our qualitative picture is unchanged if we increase artificially
the vacuum pion mass up to this value. In passing, we notice that
using a running coupling as in~\cite{Sakai:2010rp}, but adding at
the same time two further free parameters in the model, we expect
a better agreement with the Lattice. The reason is that
in~\cite{Sakai:2010rp}, the coupling $g_\sigma$ is a function of
the Polyakov loop, and it decreases as $P$ is increased.  As a
consequence, near the Polyakov loop crossover temperature, the
strength of the interaction is lowered, and a partial suppression
of the chiral condensate is expected. Quantitatively, it is not
clear a priori if the suppression is enough to rejoin the two
crossovers; only a detailed numerical study can give the answer.
We leave this important investigation to a future study.

The tiny decoupling of the two crossovers found within the PNJL
model, both in the chiral limit~\cite{Fukushima:2010fe} and in the
case of physical pion mass considered here, is observed also
within the Polyakov quark-meson model~\cite{Mizher:2010zb}, when
in the latter the zero point energy is considered (if the vacuum
energy is subtracted, then the Polyakov loop and the chiral
crossovers occur always simultaneously, but the pseudo-critical
temperature is a decreasing function of $eB$, which seems in
disagreement with the recent Lattice results~\cite{D'Elia:2010nq};
see also~\cite{Skokov:2010sf} for a recent discussion of the role
of the vacuum energy within the quark-meson model).  Since the
Polyakov loop is coupled to quarks in the same manner both in the
PNJL and in the PQM model, the tiny split of the two crossovers as
$eB$ is increased does not appear as an artifact of the PNJL
model; instead, it seems to be a consequence of the link among the
chiral condensate and the Polyakov loop, which is common in the
two kinds of models.

In the lower panels of Figures.~\ref{Fig:S3D} and~\ref{Fig:cN5},
we plot the dressed Polyakov loop as a function of temperature,
for several values of $eB$. Our definition, Eq.~\eqref{eq:SnPNJL},
differs from the canonical one~\cite{Bilgici:2008qy} for an
overall factor, which gives mass dimension one to our $\Sigma_1$.
For small values of $eB/m_\pi^2$, the behavior of $\Sigma_1$ as
temperature is increased, is qualitatively similar to that at
$eB=0$, which has been discussed within effective models
in~\cite{Kashiwa:2009ki,Mukherjee:2010cp}. In particular, the
dressed Polyakov loop is very small for temperatures below the
pseudo-critical temperature of the simultaneous crossover. Then,
it experiences a crossover in correspondence of the simultaneous
Polyakov loop and chiral condensate crossovers. It eventually
saturates at very large temperature (for example,
in~\cite{Kashiwa:2009ki} the saturation occurs at a temperature of
the order of $0.4$ GeV, in agreement with the results
of~\cite{Mukherjee:2010cp}). However, we do not push up our
numerical calculation to such high temperature, because we expect
that the effective model in that case is well beyond its range of
validity.

As we increase the value of $eB$, as noticed previously, we
observe a tiny splitting of the chiral and the Polyakov loop
crossovers. Correspondingly, the qualitative behavior of the
dressed Polyakov loop changes dramatically: the range of
temperature in which the $\Sigma_1$ crossover takes place is
enlarged, if compared to the thin temperature interval in which
the crossover takes place at the lowest value of $eB$ (compare the
solid and the dotted lines in Fig.~\ref{Fig:cN5}, as well as the
the lower panel of Fig.~\ref{Fig:S3D}).

The effective susceptibility, $d\Sigma_1/dT$, plotted in the lower
right panel of Fig.~\ref{Fig:cN5}, is qualitatively very
interesting. We observe a double peak structure, which we
interpret as the fact that the dressed Polyakov loop is capable to
feel (and hence, describe) both the crossovers. If we were to
interpret $\Sigma_1$ as the order parameter for deconfinement, and
the temperature with the largest susceptibility with the crossover
pseudo-critical temperature, then we obtain almost simultaneous
crossover even for very large magnetic field. If this were the
case, then the Polyakov loop computed within the PNJL model,
should be interpreted only as an indicator of statistical
confinement, and the deconfinement would be described by
$\Sigma_1$. Of course, this picture would not contradict the well
established picture at zero magnetic
field~\cite{Ratti:2005jh,Roessner:2006xn,Fukushima:2003fw}.
Indeed, in the case of small $eB$, we find simultaneous crossover
of chiral condensate, Polyakov loop and dressed Polyakov loop. In
the latter case, it would be just a matter of taste which quantity
one uses to identify the deconfinement crossover. Even if it is
tempting to give this kind of interpretation, which would lead to
simultaneous crossover also at finite $eB$, it is very hard to
accept it without more convincing microscopic arguments.
Therefore, in the prosecution of this work, we prefer to associate
the deconfinement crossover to that of the Polyakov loop.
Nevertheless, the dressed Polyakov loop is a new quantity which is
interesting to compute. In particular, the double peak structure
in the $\Sigma_1$ effective susceptibility, which is produced if
the magnetic field is strong enough, offers the evidence that the
dressed Polyakov loop is intimately related to both chiral
condensate and (thin) Polyakov loop, and it is capable to capture
both the crossovers. The bifurcation of the dressed Polyakov loop
susceptibility is impressive in the lower right panel of
Fig.~\ref{Fig:S3D}.

\subsection{Phase diagram in the $eB-T$ plane}

\begin{figure*}
\begin{center}
\includegraphics[width=8cm]{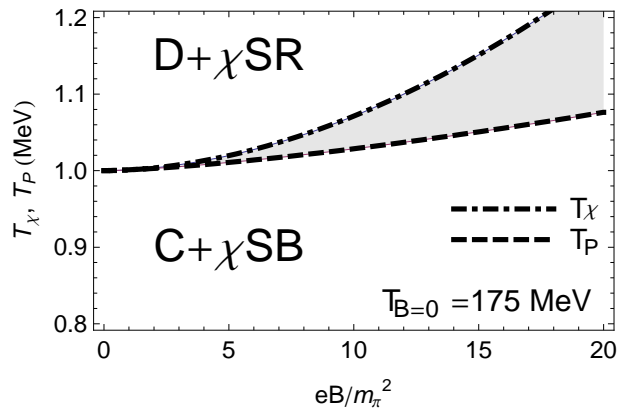}
\end{center}
\caption{\label{Fig:pd7} Phase diagram of the PNJL model in
magnetic field. Dashed line denotes the Polyakov loop crossover;
dot-dashed line corresponds to the chiral crossover. The shaded
area is the region, in the $eB-T$ plane, in which quark matter is
not statistically confined, but chiral symmetry is still broken by
the chiral condensate. Temperatures on the vertical axes are
measured in units of the pseudo-critical temperature at zero
field, which is $T_0 = 175$ MeV. The analytic form of the lines
corresponding to $T_P$ and $T_\chi$ is specified by
Eq~\eqref{eq:fit}. The UV-regulator is that corresponding to
$N=5$.}
\end{figure*}

In Figure~\ref{Fig:pd7}, we collect our results on the
pseudo-critical temperatures for chiral and Polyakov loop
crossovers, in the form of phase diagrams in the $eB-T$ plane. The
dashed line denotes the Polyakov loop crossover, and the
dot-dashed line corresponds to the chiral crossover. The shaded
area is the region, in the $eB-T$ plane, in which quark matter is
not statistically confined, but chiral symmetry is still broken by
the chiral condensate. Temperature on the vertical axes are
measured in units of the pseudo-critical temperature at zero
field, which is $T_0 = 175$ MeV. We fit our data on the
pseudo-critical temperatures by the law
\begin{equation}
\frac{T_c^{A}}{T_0} = 1 + a\left(\frac{eB}{T_0}\right)^\alpha~,
\label{eq:fit}
\end{equation}
where $A=\sigma,P$. Numerical values of the coefficients in
Eq.~\eqref{eq:fit} for the various observables are collected in
Table~\ref{Tab:fit}.  As an estimator of the goodness of the
various fits, we report in Table~\ref{Tab:fit} the percentage
error defined as
\begin{equation}
\varepsilon = 100\times\sum_{i}\left(\frac{f_A(x_i) -
y_i}{y_i}\right)^2~,
\end{equation}
where the sum runs over the data, $(x_i,y_i)$ corresponds to a
couple in the set of the data $(eB,T_A)$, and $f_A(x_i)$ denotes
the numerical value of the fit function evaluated at the data
$eB$.

\begin{table}[t!]
\caption{\label{Tab:fit}Coefficients of the fit function defined
in Eq.~\eqref{eq:fit}.}
\begin{ruledtabular}
\begin{tabular}{ccccc}
 &$a$& $\alpha$ & $T_0$ (MeV) & $\varepsilon$\\
\hline

$T_\chi$, $N=5$ &$2.4\times 10^{-3}$ &1.85 & 175 & $0.21$\\
\hline

$T_P$, $N=5$ &$2.1\times 10^{-3}$ &1.41 & 175 & $0.08$\\
\hline

$T_\chi$, $N=7$ &$7.8\times 10^{-3}$ &1.29 & 176 & 0.19 \\
\hline

$T_P$, $N=7$ &$3.9\times 10^{-3}$ &1.08 & 176 & 0.01\\

\end{tabular}
\end{ruledtabular}
\end{table}

The picture discussed in the previous Section is made clear by the
phase diagrams in Fig.~\ref{Fig:pd7}. We measure an increase of
both deconfinement and chiral crossovers; the tiny split of the
two critical temperatures is of the order of $10\%$ for the
largest value of the magnetic field strength considered here.

It is instructive to compare our results with those obtained in a
different model. The shape of the phase diagram drawn in
Fig.~\ref{Fig:pd7} is similar to that drawn by the Polyakov
extended quark-meson model, see e.g. Fig.~13 of
Ref.~\cite{Mizher:2010zb}. In that reference, an interpretation of
the split in terms of the interplay among vacuum and thermal
contribution, is given. We totally agree with those arguments,
which are reproduced within the PNJL model as well, as the final
results on critical temperatures show. In the case of the
quark-meson model, however, the picture can change even
qualitatively, depending on the fate of vacuum energy
contribution. If they are not included, then a simultaneous first
order transition is observed at every value of $eB$ (only if $eB$
is very small the transition is a smooth crossover), and the
deconfinement temperature as a function of the magnetic field
strength {\em decreases}. This picture confirms the scenario
anticipated in a previous work~\cite{Agasian:2008tb}. In the case
of the PNJL model, we cannot reproduce the latter scenario,
because of a technical reason: indeed, in our case the vacuum
contribution cannot be subtracted (as a matter of fact, we do not
have a further effective potential term at zero temperature, which
leads to spontaneous breaking of chiral symmetry when vacuum quark
contributions are subtracted). Therefore, we can limit ourselves
only to a comparison with the quark-meson model with vacuum
contributions taken into account.

\section{Conclusions}
We have computed, for the first time in the literature, the
dressed Polyakov loop for hot quark matter in external magnetic
field. To compute the finite temperature effective potential in
magnetic field, we have used the Polyakov extended Nambu-Jona
Lasinio model, with a logarithm effective action for the Polyakov
loop. In the quark sector, we have used both a four-quark and an
eight-quark interactions. Bare quark masses are fixed to reproduce
the physical value of the vacuum pion mass. This model allows to
treat self-consistently both chiral symmetry breaking and
(effective, or statistical) confinement. We improve the previous
work~\cite{Fukushima:2010fe} in three ways: we set the vacuum pion
mass to its physical value; we introduce eight-quark interaction;
finally, we compute the dressed Polyakov loop.

Our results on the dressed Polyakov loop, $\Sigma_1$, in magnetic
field show that this quantity is capable to describe both Polyakov
loop and chiral crossovers. This is resumed in the double peak
structure of the effective susceptibility $d\Sigma_1/dT$, see
Figs.\ref{Fig:S3D} and~\ref{Fig:pd7}. Moreover, we find that
$\Sigma_1$ is capable to feel both the Polyakov loop crossover and
the chiral condensate crossover, and suggests itself as the the
possibly unique order parameter of effective QCD.

The results on the pseudo-critical temperatures as a function of
$eB$ are resumed in the phase diagrams in Fig.~\ref{Fig:pd7}.
These results were anticipated in a previous
work~\cite{Fukushima:2010fe} in which only the chiral limit was
considered, and the eight quark interaction was neglected. Our
results agree qualitatively with those of
Ref.~\cite{Mizher:2010zb}, in which a quark-meson model coupled to
the Polyakov loop is considered.

As improvement of our results, it would be interesting to consider
the effects of non-locality~\cite{Hell:2008cc}. In that case,
however, the computation of the fermion spectrum in the magnetic
field is not trivial because of the non-local structure of the
action. Another interesting possibility is the use of Montecarlo
methods to compute the PNJL partition function in magnetic field,
going beyond the saddle approximation. Encouraging results along
this research line in the context of the PNJL model have been
reported in~\cite{Cristoforetti:2010sn}. Therefore, it might be
interesting to extend the computation
of~\cite{Cristoforetti:2010sn} to the case of quarks in external
magnetic field. Even more, we expect that the running coupling
introduced by the Kyushu's group~\cite{Sakai:2010rp} would help
(at least partly) to get closer crossovers in magnetic field. A
numerical investigation of this subject is left to a future study.
Finally, the extension of our calculation to finite quark chemical
potential, and to quark matter in external chromo-magnetic fields,
the latter being motivated by Lattice
results~\cite{Cea:2002wx,Cea:2007yv}, would deserve further
attention.

\acknowledgments We acknowledge correspondence with M. Huang and
S.~Nicotri, and in particular with M. d'Elia. Moreover, we
acknowledge stimulating discussions with L.~Campanelli and
K.~Fukushima. The work of M.~R.\ is supported by JSPS under the
contract number P09028. The numerical calculations were carried
out on Altix3700 BX2 at YITP in Kyoto University.


\end{document}